\documentclass[11pt,a4paper,oneside]{article}

\usepackage{jheppub}
\usepackage{color}
\usepackage{graphicx}
\usepackage{wrapfig,enumerate,slashed}

\usepackage{footmisc}
\usepackage{amsmath}
\usepackage{wasysym} 
\usepackage{graphicx}
\usepackage{subcaption}
\usepackage{color}
\usepackage{comment}
\usepackage{appendix}
\usepackage{slashed}

\newcommand{\be}{\begin{eqnarray}}
\newcommand{\ee}{\end{eqnarray}}

\newcommand{\bee}{\begin{eqnarray}}
\newcommand{\eee}{\end{eqnarray}}
\newcommand{\beeq}{\begin{equation}}
\newcommand{\eeeq}{\end{equation}}

\makeatletter
\gdef\@fpheader{}
\makeatother
\begin{document}

\title{Quantum walk approach to simulating parton showers}

\author[a]{Khadeejah Bepari,}
\author[b]{Sarah Malik,}
\author[a]{Michael Spannowsky}
\author[c]{and Simon Williams}

\affiliation[a]{Institute for Particle Physics Phenomenology, Department of Physics, Durham University, Durham DH1 3LE, United Kingdom}
\affiliation[b]{Department of Physics and Astronomy, University College London, Gower Street, London WC1E 6BT, United Kingdom}
\affiliation[c]{High Energy Physics Group, Blackett Laboratory, Imperial College, Prince Consort Road, London, SW7 2AZ, United Kingdom}

\emailAdd{khadeejah.bepari@durham.ac.uk}
\emailAdd{sarah.malik@ucl.ac.uk}
\emailAdd{michael.spannowsky@durham.ac.uk}
\emailAdd{s.williams19@imperial.ac.uk}

\abstract{This paper presents a novel quantum walk approach to simulating parton showers on a quantum computer. We demonstrate that the quantum walk paradigm offers a natural and more efficient approach to simulating parton showers on quantum devices, with the emission probabilities implemented as the coin flip for the walker, and the particle emissions to either gluons or quark pairs corresponding to the movement of the walker in two dimensions. A quantum algorithm is proposed for a simplified, toy model of a 31-step, collinear parton shower, hence significantly increasing the number of steps of the parton shower that can be simulated compared to previous quantum algorithms. Furthermore, it scales efficiently: the number of possible shower steps increases exponentially with the number of qubits, and the circuit depth grows linearly with the number of steps. Reframing the parton shower in the context of a quantum walk therefore brings dramatic improvements, and is a step towards extending the current quantum algorithms to simulate more realistic parton showers.  }

\preprint{IPPP/21/32}

\maketitle

\section{Introduction}\label{sec:intro}

The emergence of quantum computers has brought a new paradigm to the field of computation. The unique features of these devices has garnered attention from various disciplines, including High Energy Physics (HEP), where the computational challenges associated with taking, processing and analysing vast amounts of data in collider experiments like the Large Hadron Collider (LHC) requires innovative solutions. Quantum algorithms have been proposed to tackle some of these challenges, including the simulation of collision events~\cite{bauer2019quantum,PhysRevD.103.076020,li2021partonic} reconstruction of charged particle tracks in the detectors~\cite{das2020track,T_ys_z_2020,magano2021quantum}, and event classification and analysis~\cite{Blance:2020nhl,Blance:2021gcs,Terashi_2021,Wu_2021,Araz:2021zwu,Belis_2021,armenakas2020application,pires2021digital,MottQuantum}.

Collision events at the LHC typically involve hundreds of particles and can be very complicated. Simulation of such events requires extensive modelling of proton-proton interactions and the subsequent detector response to fully uncover the underlying physics processes. Theoretical descriptions of these collisions can be separated into several stages. Constituent partons in the colliding protons can interact via large momentum transfer in the so-called hard interaction. Due to the large interaction energies, such collisions have the potential to probe new physics. Colour-charged particles produced as a result of this hard interaction are likely to emit further partons, resulting in a parton shower. The parton shower process evolves the system down in energy from the hard interaction to the hadronisation scale, $\mathcal{O} ( \Lambda_{\textrm{QCD}})$. It is a perturbative process and can involve many partons, thus being one of the most time consuming parts of the generation of a collision event. Consequently, the development of quantum algorithms for the calculation of the hard process \cite{PhysRevD.103.076020} and the resultant parton shower \cite{bauer2019quantum, PhysRevD.103.076020} is an area of interest.

This paper presents a novel approach to simulating a many-particle, collinear parton shower on a quantum device using a quantum walk (QW) framework. It is structured as follows: Section~\ref{sec:QW} gives a brief introduction to the QW framework, Section~\ref{sec:PS} contains the description of the proposed parton shower algorithm, and Section~\ref{sec:Summary} gives a summary and conclusions.

\section{Quantum walks}\label{sec:QW}

The quantum random walk~\cite{PhysRevA.48.1687, Aharonov2, Kempe, Rohde2012IncreasingTD} is the quantum analogue of the classical random walk and defines the movement of a particle, the \textit{walker}, which can occupy certain $position$ states on a graph. Here we will consider only discrete-time random walks, where a coin flip determines the movement of the walker at distinct time steps. The state of the walker can therefore be defined by the position of the walker, $x$, and the coin, $c$, as $\vert x,c \rangle$. The movement of the walker through the graph is determined by two operations: the coin operation, $C$, which determines the direction the walker will move, and the shift operation, $S$, which propagates the walker to the next position. 

\begin{figure}[t]
\centering
\includegraphics[scale=0.5]{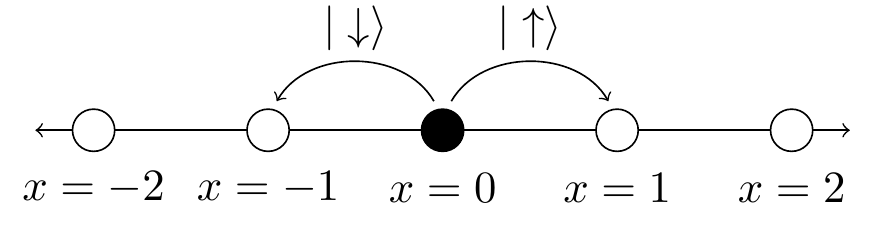}
\caption{One-dimensional walker at position $x=0$ can move either left or right depending on the outcome of the coin flip, $\vert \downarrow \rangle$ and $\vert \uparrow \rangle$ respectively.}
\label{fig:1DWalk}
\end{figure}

As a simple example, we construct a random walk following the approach in~\cite{Kempe}. Consider a walker moving along a one-dimensional line according to an unbiased coin (i.e, the walker has an equal chance of moving left or right after the coin flip), with the walker originally positioned at $x=0$, see Figure~\ref{fig:1DWalk}. The position of the particle on the line forms a Hilbert space $\mathcal{H}_P$ spanned by integer values on the line, \{$\vert i \rangle : i \in \mathbb{Z}$\}. The position space is augmented by the coin-space, $\mathcal{H}_C$, which spans two basis states, \{$\vert~\uparrow~\rangle, \vert~\downarrow~\rangle$\}, which here will represent the up and down spin-states of a fermion\footnote{The coin space can be represented by any two-level quantum system. The choice of using the up and down spin-states of a fermion is useful when implementing quantum walks on qubit-based quantum devices, such as those available on the IBM Q network \cite{IBMQ}.}. Therefore, the walker occupies a total space of

\begin{equation}
\mathcal{H} = \mathcal{H}_C \otimes \mathcal{H}_P.
\end{equation}
In the classical case, the coin operation is carried out by evaluating a classical coin. Based on the outcome of this coin, the shift operator moves the walker in the correct direction. Here we will attribute the coin state $\vert \uparrow \rangle$ to the walker moving in the positive $x$ direction and the $\vert \downarrow \rangle$ state to the walker moving in the negative $x$ direction. After the step process is complete, the walker is either in the $x=-1$ or $x=1$ position. In contrast to the classical case, the quantum coin operation is based on a \textit{quantum coin}. In this example, we will consider the Hadamard coin,

\begin{equation}
H = \frac{1}{\sqrt{2}} \begin{pmatrix} 1 & \phantom{-}1 \\ 1 & -1 \end{pmatrix},
\end{equation}
which gives an equal chance for the coin qubit to be measured in each of the coin states. The quantum coin operation puts the system into a superposition of the basis states of the $\mathcal{H}_C$ space. The shift operation is then performed, moving the walker into a superposition of the position states, $x=-1$ and $x=1$. If a measurement is performed after the step, the wavefunction collapses to recover the classical case of the walker being in either the $x=-1$ or $x=1$ position.

\begin{figure}[t]
\centering
\includegraphics[scale=0.4]{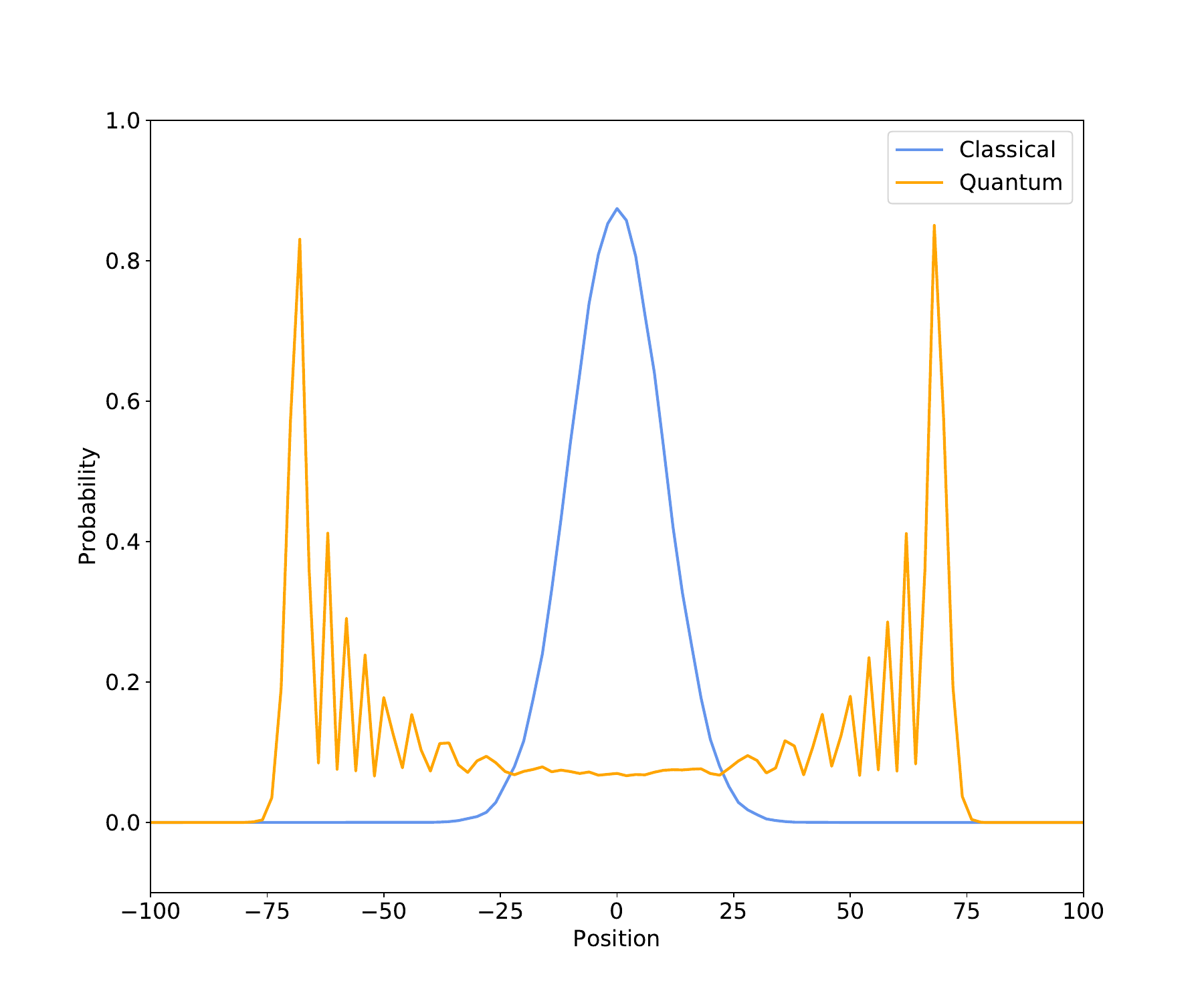}
\caption{Simulation of a 100-step random walk using the IBM Q 32-qubit simulator~\cite{32_sim} for 100,000 shots for a classical random walk obtained by measuring the coin state after each step, and  a quantum random walk using a symmetric initial position and a Hadamard coin. Only non-zero probabilities are shown, as odd-numbered positions will have zero probability for this walk.}
\label{fig:100StepWalks}
\end{figure}

The Hadamard coin used here is a balanced unitary coin operation\footnote{Strictly speaking, the Hadamard coin introduces a bias to the quantum walk through the phase on the coin qubit. This is discussed in detail in \cite{Kempe} and references therein. Here we remove this bias for a general quantum walk by using a symmetric initial state, \begin{equation} \vert \Phi_\textrm{symm} \rangle = \frac{1}{\sqrt{2}} \big(\vert 0 \rangle + i \vert 1 \rangle \big) \otimes \vert 0 \rangle,\end{equation} obtaining the distribution presented in Figure~\ref{fig:100StepWalks}.} and therefore the coin and shift operations can be defined as a single unitary transformation to the initial qubit state,

\begin{equation}\label{eqn:QWprocess}
U = S \cdot (C \otimes I),
\end{equation}
which is applied iteratively to represent the number of steps. For a quantum walk of $N$ steps, the propagation of the walker is described by the transformation $U^N$ \cite{Kempe}. An example of running a linear, one-dimensional, $N=100$ step, random walk for both the classical case and the quantum case is shown in Figure~\ref{fig:100StepWalks}. The classical case has been achieved by measuring the coin qubit at each step, removing the superposition from the system. As expected, the classical walk yields a Gaussian distribution of positions centred about the initial position of the particle, with the variance $\sigma^2 = N$. In contrast, the quantum random walk, where there are quantum interference effects between the intermediate steps of the walk process, results in a distribution that is dramatically different from the classical case. It can be shown \cite{Kempe, Ambainis} that the variance of the quantum random walk process goes as $\sigma^2 \sim N^2$. This is a remarkable attribute of the quantum random walker, which propagates quadratically faster than the classical walker. The average distance of the walker from the initial position is $\sigma = \sqrt{N}$ and $\sigma \sim N$ for the classical and quantum walks respectively.

\subsection{Quantum Walks with Memory}\label{sec:QWM}

An interesting extension of the discrete-time quantum walk is the addition of a ``\textit{memory}" Hilbert space. Analogous to classical random walks with memory, quantum walks with memory offer an opportunity to simulate arbitrary dynamics by modifying the movement of the walker based on the outcomes of previous coin operations~\cite{Brun, Rohde} and walker positions~\cite{Camilleri}. Through the use of memory, quantum walks have been shown to display unique diffusive characteristics~\cite{Brun, Rohde, Camilleri, Li2016}. The diffusive characteristics depend on the size of the quantum walk's memory, and the probability distributions range from an ideal quantum walk distribution, to an ideal classical walk distribution, in the limit of full memory~\cite{Brun, Rohde}. 

An example of a discrete-time quantum walk with memory is presented in~\cite{Rohde} and considers quantum walks with different size memory of previous coin outcomes. As the size of the quantum memory increases to the number of steps in the walk, this is equivalent to a new coin per step and classical statistics are obtained. Classical distributions can be easily obtained by measuring the coin, or introducing decoherence, at each step. Therefore, one might assume that, in the limit of full memory, quantum walks are reduced to classical algorithms. However, and very importantly, there is no decoherence present within the quantum walk with memory and the state is still a highly entangled pure state~\cite{Rohde}. The evolution of the quantum walk is entirely unitary\footnote{It is worth noting that the construction of a unitary evolution for a quantum walk with memory is not trivial, and a generic model for all quantum walks with memory is yet to be outlined. Reference~\cite{Li2016} outlines a generic model for all quantum walks with memory on regular graphs.}, thus one can always evolve back to the initial state by reversing time. This is not possible in a truly classical case. Furthermore, quantum walks with memory have been shown to spread significantly faster than their classical counterparts~\cite{Camilleri}.

The ability of quantum walks with memory to efficiently simulate classical dynamics through the population of a memory Hilbert space is particularly useful in the simulation of parton showers on quantum devices. Section~\ref{sec:PS} outlines the implementation of a discrete collinear QCD parton shower on a quantum device using a quantum walk with memory of the previous step's coin operation.

\section{Quantum walk as a parton shower simulation}\label{sec:PS}

The parton shower \cite{Buckley:2011ms} evolves the energy scale of a scattering event from the hard interaction down to the hadronisation scale through the radiation of additional partons. The emissions are determined by splitting functions which correspond to the different emission probabilities in the shower. The shower content is then updated depending on which splitting probability is chosen. In classical computing, parton shower algorithms are implemented using Markov Chain Monte Carlo (MCMC) algorithms~\cite{Herwig, Sherpa, Pythia} to efficiently sample the probability distributions of the shower final state observables. In quantum computing, a quantum walk mechanism provides a natural framework for the simulation of this probabilistic interpretation of parton showers: the emission probabilities correspond to the coin flip probabilities, and updating the shower content depending on the emission corresponds to the shift operation in the quantum walk framework.

In this Section, we detail this novel quantum walk approach to simulating a parton shower on a quantum device. Within this framework, the algorithm can simulate a many-particle parton shower, and shows a remarkable improvement on the number of shower steps that can be simulated in comparison to previous quantum algorithms~\cite{bauer2019quantum, PhysRevD.103.076020}. The Section is ordered as follows: Section~\ref{sec:Theory} gives the theoretical outline of the toy model used in the parton shower, Section~\ref{sec:SimpleShower} shows the implementation of a simple parton shower with one particle type, Section~\ref{sec:collinearShower} outlines the full collinear parton shower, Section~\ref{sec:speedup} examines the quadratic speed up provided by a quantum walk under certain conditions and highlights parallels with the quantum walk parton shower, and Section~\ref{sec:future} discusses possible extensions to the algorithm with advancements in quantum computers to simulate a realistic parton shower.

\subsection{Theoretical outline of the shower algorithm}\label{sec:Theory}

We present a discrete, collinear parton shower using the quantum walk framework. As with the parton shower algorithms presented in~\cite{PhysRevD.103.076020, bauer2019quantum}, this algorithm utilises the ability of the quantum device to remain in a superposition state throughout the calculation. Consequently, all shower histories are calculated simultaneously and are encoded in the final wavefunction, with a measurement projecting out a specific quantity of the final state, e.g. the number of partons. This offers a unique advantage over classical parton shower algorithms, which need to calculate each shower history explicitly and store the information on a physical memory device. Only after summing over all possible shower histories can a physically meaningful quantity be extracted. The goal of this algorithm is to create the foundation for the development of a full, general parton shower by studying a simplified toy model that meets the capabilities of current quantum simulators.

An emission is collinear if a parton with momentum $P$ splits into two massless particles, which have parallel 4-momenta, such that the momentum distribution is, 

\begin{align}
p_i = zP,& &p_j = (1-z)P,
\end{align}
thus, $(p_i + p_j)^2 = P^2 = 0$ \cite{Taylor_2017}. In this algorithm, we use a similar theoretical set-up as~\cite{PhysRevD.103.076020}. In each shower step, emission is determined by first ascertaining whether an emission occurred in the step using the Sudakov factors, and then applying the relevant splitting functions. The Sudakov factors for a QCD process are given by, 

\begin{align}
\Delta_{i,k}(z_1, z_2) = \exp \big[ - \alpha_s \int^{z_2}_{z_1} P_k (z^\prime) dz^\prime \big],
\end{align}
and are used to calculate the probability of non-emission~\cite{Sudakov:1954sw}. The probability that no particles split for an arbitrary step $N$ in the shower process, where $N$ particles can be present, is given by the total Sudakov factor, 

\begin{equation}\label{eqn:Sudakovs}
\Delta_\textrm{tot} (z_1,z_2) = \Delta_g^{n_g}(z_1,z_2)\Delta_q^{n_q}(z_1,z_2)\Delta_{\overline{q}}^{n_{\overline{q}}}(z_1,z_2)
\end{equation}
where $n_g$, $n_q$ and $n_{\overline{q}}$ are the number of gluons, quarks and antiquarks present in the step\footnote{As the algorithm allows for steps with no emissions, for a step $N$: $(n_g + n_q + n_{\overline{q}}) \leq N$}. 
As in~\cite{PhysRevD.103.076020}, only collinear splittings will be considered. The emission probabilities are therefore calculated using the collinear splitting functions outlined in \cite{Dokshitzer:1977sg, Gribov:1972ri, ALTARELLI1977298}. The emission of a gluon from a quark is defined at leading order (LO) by,

\begin{align}\label{eqn:QSplit}
P_{q\rightarrow qg} (z) = C_F \frac{1 + (1-z)^2}{z},
\end{align}
where $C_F = 4/3$ is calculated using colour algebra, and the quark and gluon have momentum fractions $(1-z)$ and $z$ respectively. The gluon can self-couple, and therefore can split to both a pair of gluons and a quark-antiquark pair. At LO, the splitting functions for these emissions are, 

\begin{align}\label{eqn:GSplit}
P_{g\rightarrow gg} (z) = C_A \Big[ 2 \frac{1-z}{z} + z(1-z) \Big],& &P_{g\rightarrow q\overline{q}} (z) = n_f T_R (z^2 + (1 - z)^2),
\end{align}
where $C_A = 3$ and $T_R = 1/2$ are calculated using colour algebra, and $n_f$ is the number of massless quark flavours.

Combining the Sudakov factors with the splitting functions defines the full probability of emission for particle $k$ splitting to $i$ and $j$,

\begin{equation}\label{eqn:totalProb}
\textrm{Prob}_{k\rightarrow ij} = (1 - \Delta_k) \times P_{k \rightarrow ij} (z). 
\end{equation}
In the QW framework, this probability is applied as a unitary rotation to the coin qubit, defining the shower algorithm's coin operation.

The proposed algorithm does not include kinematics. This allows for the calculation to be implemented on currently accessible simulators, such as the 32-qubit IBM Q Quantum Simulator~\cite{32_sim}. As a result, the shower evolution cannot be determined by the kinematics of the shower particles, but instead the evolution variable $z$ is evolved to lower and lower momenta, exponentially, with the number of steps. Section~\ref{sec:future} outlines how a more realistic parton shower could be constructed on future devices.

\subsection{Implementation of a simple shower as a one-dimensional quantum walk}\label{sec:SimpleShower}

The implementation of a parton shower as a quantum walk follows the framework of a simple quantum random walk outlined in Section~\ref{sec:QW}. Here we define the coin operation as a unitary rotation on the coin qubit corresponding to the probability of emission, calculated using the Sudakov factors and the subsequent splitting functions defined in Section~\ref{sec:Theory}. This rotation takes the form,

\begin{equation}\label{eqn:coin}
U_c = \begin{pmatrix} \sqrt{1 - P_{ij}} & -\sqrt{P_{ij}} \\ \sqrt{P_{ij}} & \phantom{-}\sqrt{1 - P_{ij}} \end{pmatrix},
\end{equation}
where $P_{ij}$ is the probability of particle $k$ splitting to $i$ and $j$, as defined in Equation~\ref{eqn:totalProb}. The coin space, $\mathcal{H}_C$, therefore spans the space \{$\vert 0 \rangle$, $\vert 1 \rangle$\} defined by the possible measured states of the coin qubit. Here we define the $\vert 0 \rangle$ state as the ``\textit{no emission}" state, and the $\vert 1 \rangle$ state as the ``\textit{emission}" state. The position space, $\mathcal{H}_P$, now defines the number of particles present in the shower and has been altered to include only zero and positive integers, \{$\vert i \rangle : i \in \mathbb{N}_0$\}, as the parton shower cannot have a negative number of particles. The shift operation is controlled from the coin qubit and moves the walker in the correct direction. 

\begin{figure}[t]
\centering
\includegraphics[width=1\textwidth]{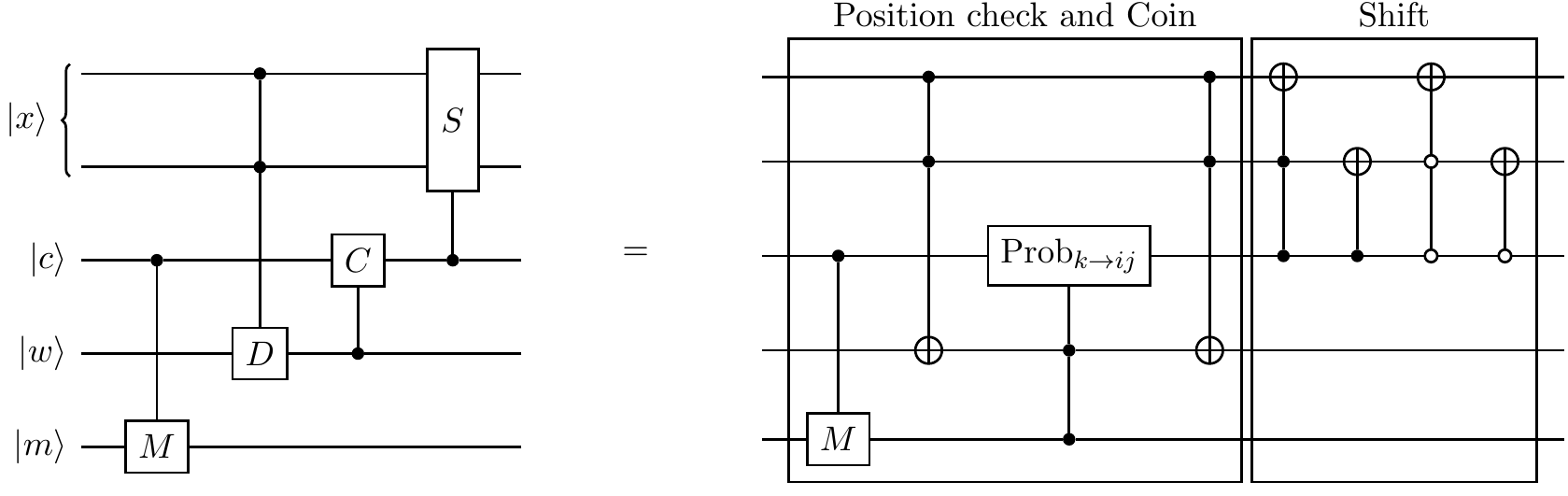}
\caption{Schematic for a single step of a quantum walk as a parton shower, with the ability to simulate a particle which can split to more particles of the same type. Here, the ``\textit{position check}" determines the number of particles present by assessing the position of the walker and records the state of the coin qubit to a memory register (only after the second step). The ``\textit{coin}" operation applies the correct splitting probabilities depending on the position of the walker. The ``\textit{shift}" operation moves the walker depending on the outcome of the coin operation. Here only one position possibility is shown. The ``\textit{position check}" will loop through all possible walker positions to correctly apply the coin operation.}
\label{fig:1DShower}
\end{figure}

To illustrate this simple shower, Figure~\ref{fig:1DShower} shows a schematic of a one-dimensional quantum walk with memory able to simulate a particle which can split to produce another particle of the same type. In this simple shower, the number of particles present is encoded in the position of the walker, with the initial state of the walker being the zero position. Figure~\ref{fig:1DShower} uses a two qubit basis for the position of the walker, ultimately allowing the algorithm to simulate a maximum of 4 shower particles in the final state. The number of particles that the algorithm can simulate increases exponentially with the number of position qubits, $x$, as $2^x$. For this example, only one splitting is possible, $i \rightarrow ii$, and as a result only one coin qubit is needed to encode the splitting probability. As outlined in Section~\ref{sec:Theory}, the splitting probability from Equation~\ref{eqn:totalProb} contains the Sudakov factor. To correctly implement the splitting probability for the shower step, the particle content of the shower must be known. This is handled by the ``\textit{position check}" scheme illustrated in Figure~\ref{fig:1DShower}, which controls from the position of the walker and applies the correct splitting probability accordingly. The scheme is constructed from a series of \textsc{ccnot} gates, thus the operation is entirely unitary and maintains the coherence of the system. Futhermore, the ``\textit{position check}" scheme ensures that the coin operation from Eqn.~\ref{eqn:coin} is always applied to the $\vert 0 \rangle$ state on the coin qubit to recover the correct parton shower distribution. This is done by allowing the walker to have a memory of the outcome of the coin from the previous step by populating a memory register. For the quantum walk parton shower, memory is only required after the second shower step and thus the algorithm does not have full memory of the walk evolution. Quantum walks with memory are discussed in Section~\ref{sec:QWM}.

The subsequent shift operation then adjusts the number of particles present in the shower, depending on the outcome of the coin operation. If, after the coin operation, the coin qubit is in the $\vert 1 \rangle$ state, then the splitting has occurred and the position of the walker is increased by one, increasing the number of particles present in the shower by one. However, if the coin operation yields a $\vert 0 \rangle$ state, then the walker does not move for this simple example\footnote{Note that in Figure~\ref{fig:1DShower} the shift operation also shows the ability to decrease the walker's position. This is not needed for the simple example of $i\rightarrow ii$ splittings, but will be useful later.}. As shown in Figure~\ref{fig:1DShower}, the shift operation is constructed from a series of Toffoli gates and thus is a unitary operation.

This step can then be repeated for the number of discrete shower steps in the parton shower, resembling the quantum random walk outlined in Section~\ref{sec:QW}. Throughout the calculation, the device is in a superposition state of all possible outcomes of the coin and shift operations. At the end of the shower process, the final state of the system is measured and projected onto a classical state.

\subsection{Implementation of a collinear parton shower}\label{sec:collinearShower}

It is possible to extend the simple shower outlined in Section~\ref{sec:SimpleShower} to include multiple parton types by increasing the dimension of the position space $\mathcal{H}_P$, with the aim of developing a multi-particle, discrete, collinear parton shower using the theoretical outline discussed in Section~\ref{sec:Theory}. The algorithm presented here considers a toy model comprised of a gluon and one flavour of quark, and can simulate the corresponding splittings.

\begin{figure}[t]
\centering
\begin{subfigure}{0.49\textwidth}
\centering
\includegraphics[width=0.9\textwidth]{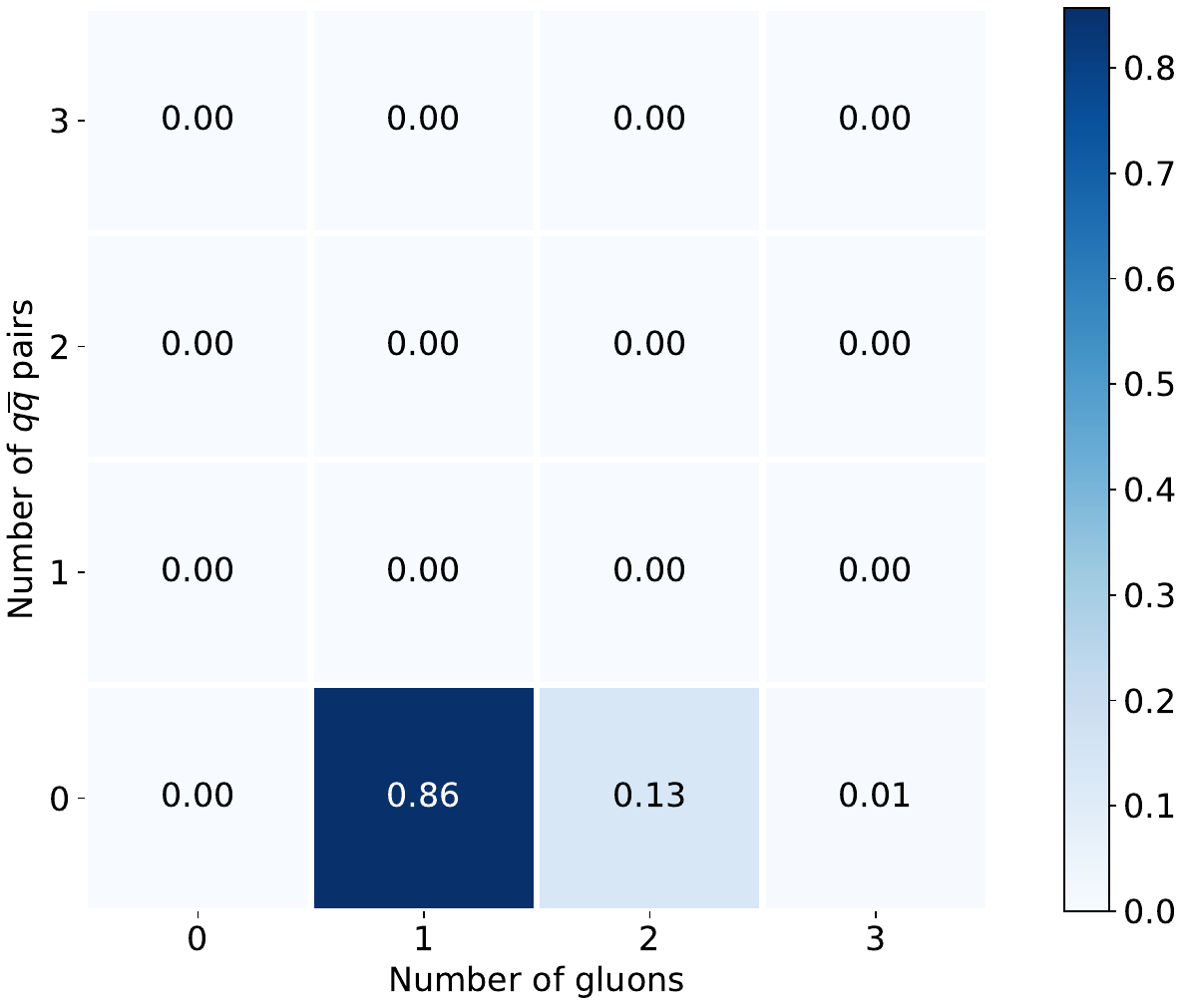}
\subcaption{}
\end{subfigure}
\begin{subfigure}{0.49\textwidth}
\centering
\includegraphics[width=0.9\textwidth]{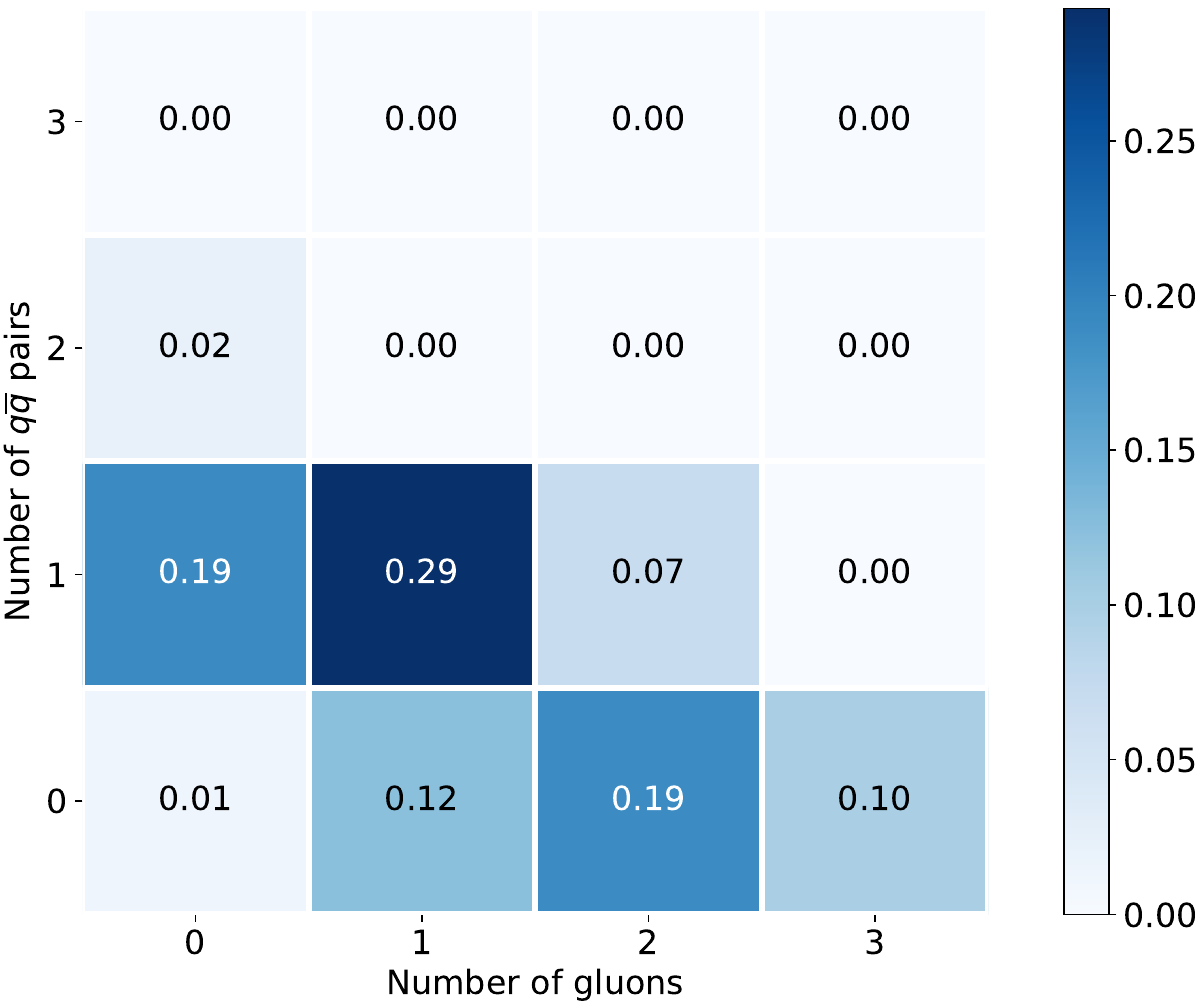}
\subcaption{}
\end{subfigure}
\caption{Visualisation of a quantum walk as a parton shower comprising gluons and quarks. The quantum walker's position on a 2D plot corresponds to the number of particles in the parton shower: (a) shows a parton shower using the collinear splitting functions for quarks and gluons, (b) shows a parton shower with modified splitting functions to show how the walker moves in the 2D lattice.}
\label{fig:visualisation}
\end{figure}

\begin{figure}[t]
\centering
\includegraphics[scale=0.3]{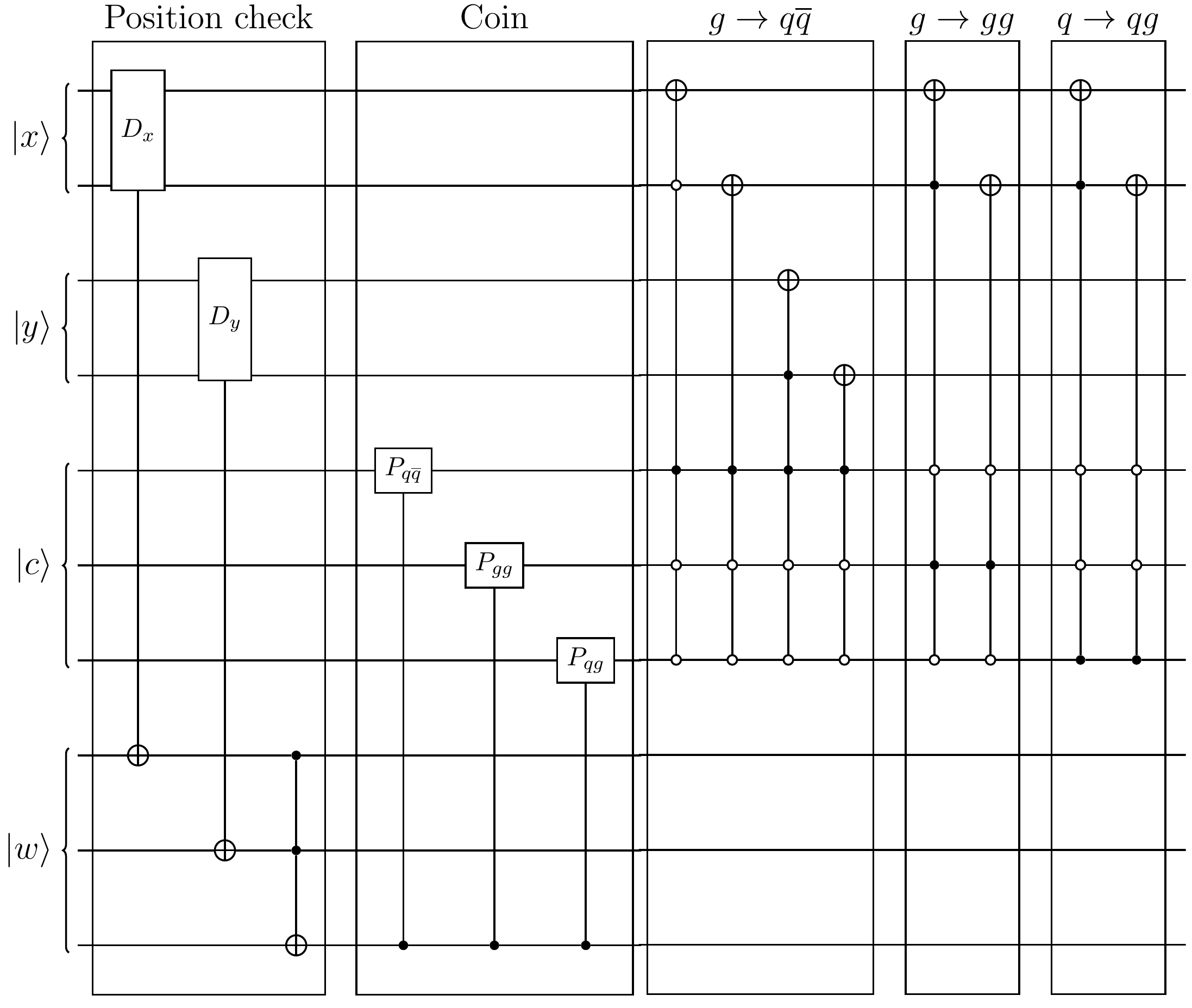}
\caption{Schematic of the quantum circuit for a single step of a discrete QCD, collinear parton shower with the ability to simulate the splittings of gluons and one flavour of quark. The shower algorithm is split into three distinct sections: (1) The position check determines the position of the walker so that the correct Sudakov form factors are applied in the splitting kernels, (2) the coin flip applies unitary rotations to a coin register corresponding to the possible splitting kernels, (3) the shift operation propagates the walker into the correct direction to describe the particle splitting in the shower step. This step is then repeated iteratively to simulate a full shower process.}
\label{fig:partonShower}
\end{figure}

As shown in Section~\ref{sec:SimpleShower}, a quantum walker in a one-dimensional position space, $\mathcal{H}_P$, has the ability to simulate a single particle type. Augmented by the coin space, $\mathcal{H}_C$, with dimension equal to the number of possible splittings associated with the particle, the quantum walk can simulate a simple parton shower comprising one particle type. Increasing the dimension of the position space increases the number of particles which can be simulated in the algorithm. Applying this mechanism to our toy model of the parton shower, the position space, $\mathcal{H}_P$, is increased to two dimensions to accommodate the simulation of gluons and quarks, counting gluons in one dimension and quarks in the other. Note that we do not need to include dimensions for both quarks and antiquarks as they are produced in conjunction through the $g\rightarrow q\overline{q}$ splitting, thus instead we count quark-antiquark pairs. Figure~\ref{fig:visualisation} shows a visualisation of how the walker's position on a 2D plot corresponds to the number of particles in the shower, with gluons and quarks measured on the $x$ and $y$-axes respectively of the walker's 2D lattice. The coin space $\mathcal{H}_C$ is increased to a three-dimensional Hilbert space, with three coin qubit rotations corresponding to the splitting functions in Equations~\ref{eqn:QSplit} and \ref{eqn:GSplit}. Controlled from the coin register, the shift operations propagate the walker to reflect the production of new particles in the shower step. A schematic of the quantum circuit is shown in Figure~\ref{fig:partonShower}. It should be noted that it is likely that more than one of the coin qubits can be in the $\vert 1 \rangle$ state in a step. In these situations, it is not clear which splitting kernel should be applied and therefore the algorithm does not apply a shift operation to the walker. This is realised by controlling from coin states that only have one coin qubit in the $\vert 1 \rangle$, as shown in Figure~\ref{fig:partonShower}.

To simulate a parton shower, the step shown in Figure~\ref{fig:partonShower} is performed many times. The qubit requirement for an $N$-step shower scales as

\begin{equation}\label{eqn:scaling}
n_\textrm{qubits} = 2 \log_2 (N+1) + 6.
\end{equation}
Only one splitting is allowed to occur per step, and steps where no emission occurred are dictated by the Sudakov form factors from Equation~\ref{eqn:Sudakovs}. The system is kept in a superposition state throughout the algorithm, with a measurement taking place only at the end of the calculation. Therefore, after all the steps have been evaluated, the system is in a superposition of all possible shower histories. This differs dramatically from classical parton shower algorithms where each shower history must be individually calculated. A physically meaningful quantity can only be extracted from a classical shower algorithm once all possible shower histories have been summed over. Consequently, the quantum algorithm approach to parton showers provides a unique advantage over the classical approach. 

\begin{figure}[t]
\centering
\includegraphics[width=\textwidth]{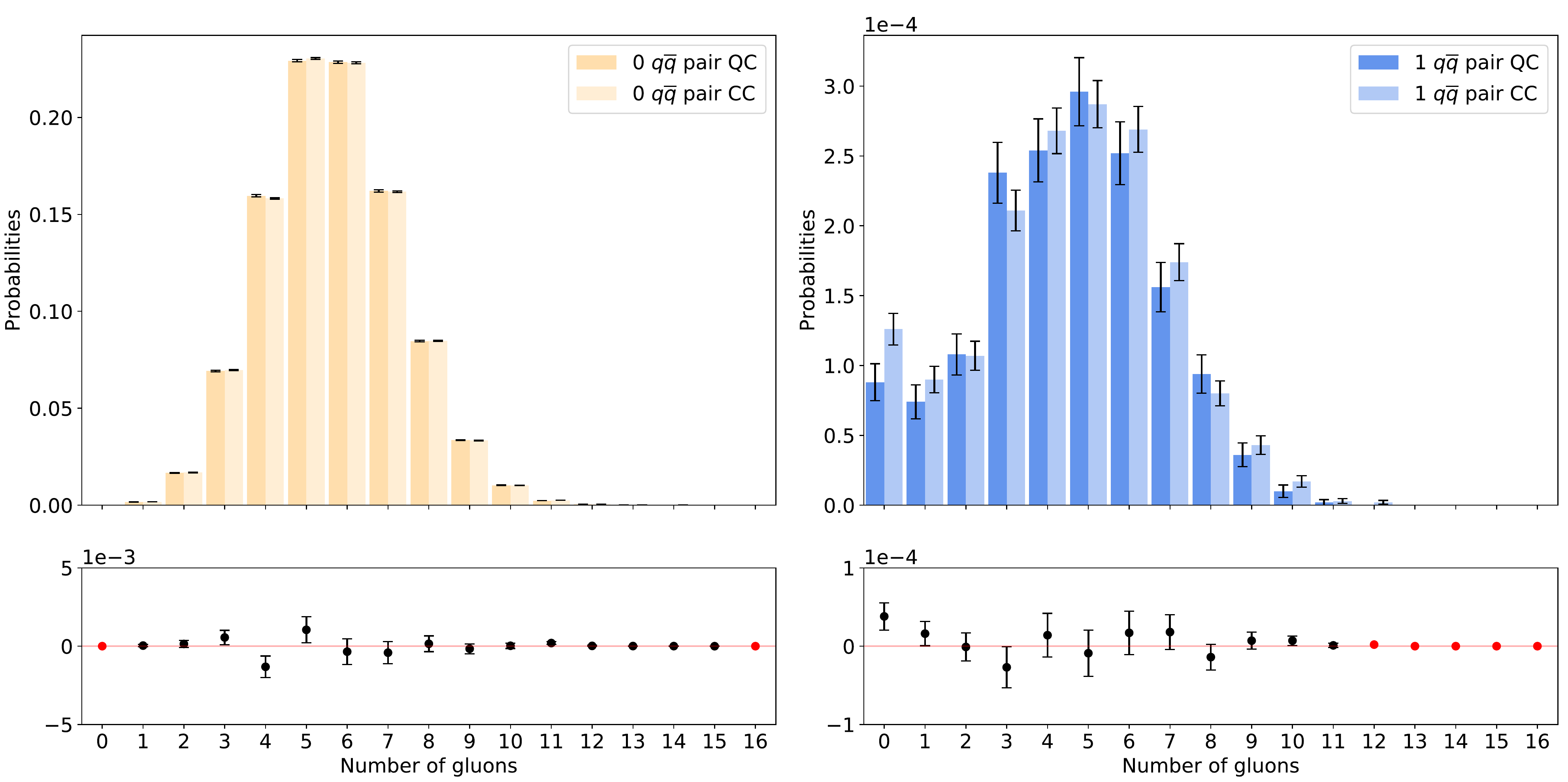}
\caption{Probability distribution of the number of gluons measured at the end of the 31-step parton shower for the classical and quantum algorithms, for the scenario where there are zero quark-antiquark pairs (left) and exactly one quark-antiquark pair (right) in the final state. The quantum algorithm has been run on the IBMQ 32-qubit quantum simulator~\cite{32_sim} for 500,000 shots, and the classical algorithm has been run for $10^6$ shots.}
\label{fig:originalResults}
\end{figure}

\begin{figure}[t]
\centering
\includegraphics[width=\textwidth]{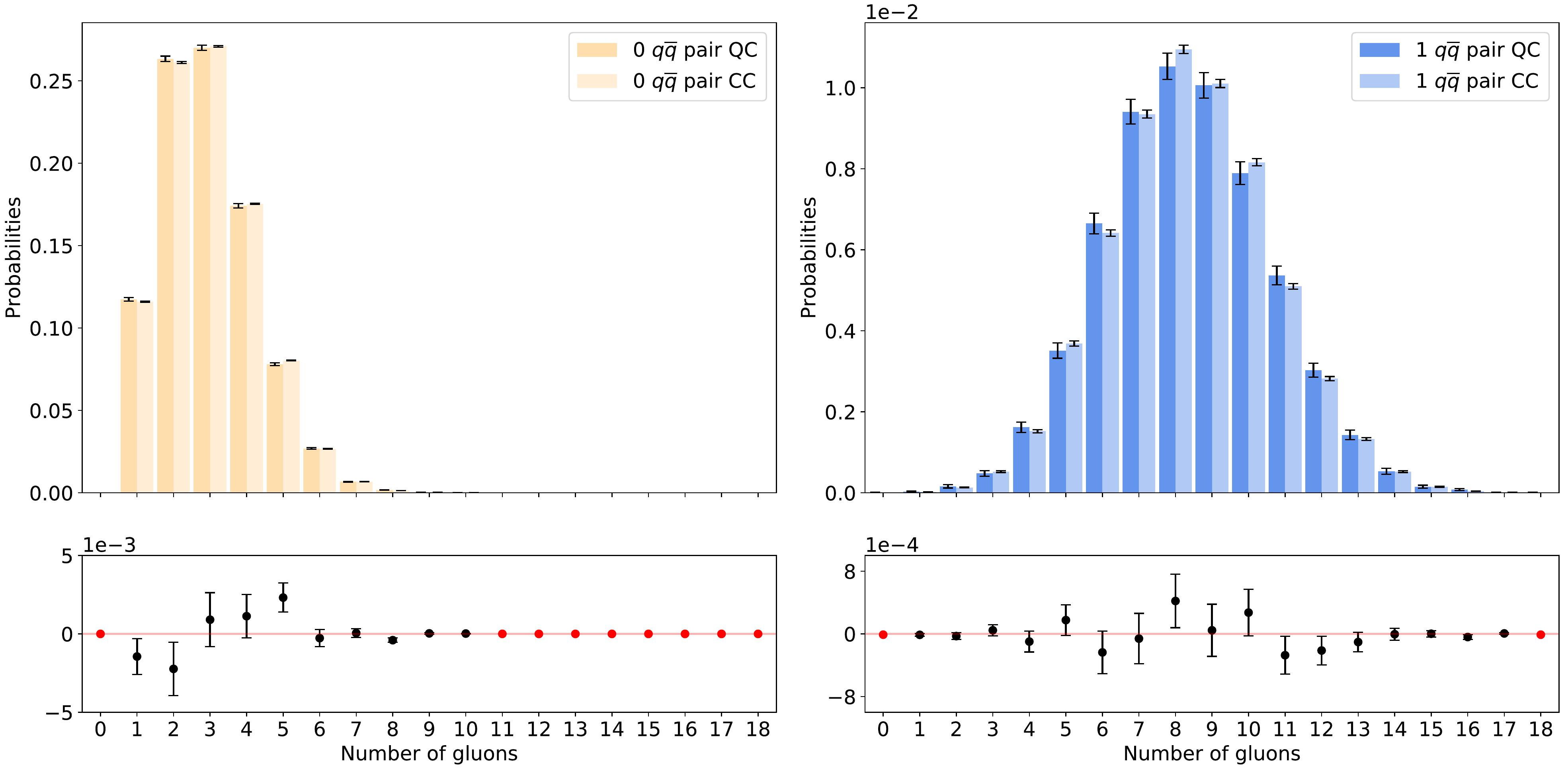}
\caption{Probability distribution of the number of gluons measured at the end of the 31-step parton shower for the classical and quantum algorithms with modified splitting kernels, for the scenario where there are zero quark-antiquark pairs (left) and exactly one quark-antiquark pair (right) in the final state. The quantum algorithm has been run on the IBMQ 32-qubit quantum simulator~\cite{32_sim} for 100,000 shots, and the classical algorithm has been run for $10^6$ shots.}
\label{fig:modResults}
\end{figure}

The quantum parton shower algorithm with 31 shower steps has been run for 500,000 shots on the IBM Q 32-qubit Quantum Simulator~\cite{32_sim}. The output from the quantum simulator has been compared to a classical parton shower algorithm, which follows the same theoretical framework as that outlined in Section~\ref{sec:Theory}, simulating a toy model with one quark flavour and a gluon. Figure~\ref{fig:originalResults} shows the comparison between the probability distributions produced by the quantum and classical parton shower algorithms for the number of gluons measured at the end of the shower. This is shown for the scenario where there are zero quark-antiquark pairs in the final state and the much less probable scenario where there is one quark-antiquark pair in the final state. Due to the low statistics for the $1 q\overline{q}$ results, a further validation of the parton shower algorithm has been carried out using modified splitting functions to enhance the $g \rightarrow q\overline{q}$ and $q \rightarrow qg$ splittings. The results of this test are shown in Figure~\ref{fig:modResults} and display good agreement between the quantum and classical parton shower algorithms. The probability of producing two quark-antiquark pairs is less than $10^{-5}$. For both comparison runs, the classical algorithm has been executed for 31 shower steps with $10^6$ shots of the algorithm.

The quantum walk framework provides a distinct increase in the qubit and circuit depth efficiency of the parton shower in comparison to previous quantum parton shower algorithms~\cite{bauer2019quantum, PhysRevD.103.076020}. As shown in Figure~\ref{fig:originalResults}, the quantum walk parton shower has the ability to simulate over 15 times as many shower steps than~\cite{PhysRevD.103.076020}, and requires just over half the amount of qubits. As a direct comparison, the quantum walk parton shower can recreate the 2-step shower presented in~\cite{PhysRevD.103.076020} using 9 qubits and 203 gate operations (59 single qubit gates, 98 \textsc{ccnot} gates and 46 \textsc{cnot} gates); a marked improvement on the 31 qubits and 444 gate operations (169 single qubit gates, 217 \textsc{ccnot} gates and 58 \textsc{cnot} gates) required in~\cite{PhysRevD.103.076020}. Furthermore, the quantum walk parton shower scales much more efficiently than the shower in~\cite{PhysRevD.103.076020}. The number of partons that can be simulated in the final state grows exponentially with the number of qubits in the quark and gluon registers. Due to the efficiency of the quantum walk framework, the circuit depth grows linearly with the number of shower steps. In contrast, the number of qubits required to perform the history operation in~\cite{PhysRevD.103.076020} grows, at best, quadratically with the number of partons in the final state. The proposed quantum walk parton shower algorithm therefore provides a significant enhancement in a quantum device's ability to simulate a realistic parton shower. 

It is natural to consider whether quantum computers and the quantum walk framework provide an advantage for simulating parton showers over classical algorithms. Further to the arguments made here and in \cite{PhysRevD.103.076020}, it is interesting to examine the possible speed up provided by quantum walks for parton shower algorithms.

\subsection{Quantum walk speed up of classical Markov Chain Monte Carlo algorithms}\label{sec:speedup}

In classical computation, parton showers are simulated using Markov Chain Monte Carlo (MCMC) algorithms~\cite{Herwig, Sherpa, Pythia} to efficiently sample from the probability distribution of shower final state observables. Beyond parton shower applications, MCMC algorithms are the heart of modern computation, with applications across many fields. Consequently, there is much discussion in the quantum information community as to whether quantum computers can provide a speed up in comparison to classical MCMC techniques~\cite{Montanaro}. Quantum walks have been shown to give quadratic speed up for certain cases such as search algorithms~\cite{PhysRevA.67.052307, Aharonov2, Szegedy, doi:10.1137/090745854}, simulated annealing~\cite{PhysRevA.78.042336, PhysRevLett.101.130504} and dedicated circuits for MCMC algorithms~\cite{Lemieux2020efficientquantum}. The important factor that leads to the speed up compared to these MCMC algorithms comes from the Markov Chain mixing time, the time it takes for the algorithm to approach the equilibrium distribution~\cite{levin2008markov, sinclair1993algorithms}. The mixing time of the Markov Chain is therefore directly linked to the run time of the algorithm. It is interesting to note that classical MCMC algorithms are not optimal in general.

It can be shown that, for a stochastic matrix $M$ describing a random walk, the mixing time is related to the so called ``spectral gap" between the largest eigenvalue of the stochastic process and the second largest eigenvalue~\cite{Kempe, Aharonov2, Lemieux2020efficientquantum}. For the minimal spectral gap $\delta$, the mixing time of the classical random walk is known to be proportional to $\delta^{-1}$. The spectral gap for the quantum walk is quadratically larger in comparison to its classical analogue, therefore the mixing time is decreased quadratically. In combination with an efficient implementation time of a single step, this results in an algorithm that reaches the equilibrium distribution quadratically faster than a classical MCMC~\cite{Lemieux2020efficientquantum, PhysRevA.78.042336, PhysRevLett.101.130504}. Although it is not known whether this quadratic speed up can be obtained in general for all MCMC algorithms~\cite{PhysRevA.76.042306, Orsucci2018}, this is still an active area of research with upper bounds on the mixing times of quantum walk Markov Chain algorithms recently being estimated~\cite{PhysRevLett.124.050501, PhysRevA.104.032215}.

Of particular interest to the parton shower algorithm presented here is the quantisation of ergodic Markov Chains using quantum walk methods, such as the Szegedy quantisation~\cite{Szegedy}, and equivalent coin based~\cite{WongCoinedWalk, Portugal} and memory~\cite{Li2020} quantum walks. It has been proven that a quadratic speed up in mixing time can be achieved for algorithms based on the implementation of reversible, ergodic Markov Chains~\cite{doi:10.1137/090745854, Lemieux2020efficientquantum}. A Markov Chain is said to be ergodic if it is aperiodic and irreducible, and consequently has a unique equilibrium distribution~\cite{norris_1997}.

\subsubsection{Potential speed up of the quantum walk parton shower}

There are many parallels between the quantum walk parton shower and the quantum walks discussed in Section~\ref{sec:speedup}, which have a proven speed up over classical MCMC algorithms. To highlight this, we consider the observable for the expected number of gluons $\langle n_g \rangle$, represented by the walker's position along the $x$-axis in Figure~\ref{fig:visualisation}. The shower is initiated in the $\langle n_g = 1\rangle$ state, with the walker in the $(x,y)=(1,0)$ position on the two dimensional lattice. At each step, the possible splittings mean that the number of gluons can be increased, by the $g\rightarrow gg$ and $q\rightarrow qg$ splittings, and decreased, by the $g\rightarrow q\overline{q}$ splitting. Consequently, the evolution of the observable $\langle n_g \rangle$ can be depicted as a binary tree walk, with the walker having the ability to move both left and right on the $x$-axis to any state in the $\langle n_g \rangle$ state space after some number of steps, $N$. Therefore, the Markov Chain representing the observable $\langle n_g \rangle$ is irreducible as there are no transient states in the number of gluons. Furthermore, the parton shower is aperiodic and thus $\langle n_g \rangle$ is ergodic when $g\rightarrow q\overline{q}$ splittings are included.  

Therefore, there is a connection between the quantum walk parton shower, and algorithms which achieve quadratic speed up for sampling ergodic Markov Chains~\cite{Szegedy, doi:10.1137/090745854, Lemieux2020efficientquantum}. As a consequence, the quantum walk parton shower has the potential to benefit from a quadratic speed up in sampling observables that satisfy these conditions, such as the number of gluons, $\langle n_g \rangle$ in the final state. This article does not consider an analytical determination of potential speed up and focuses on improving the required Quantum Volume compared to previous quantum shower algorithms, leading to more realistic shower depths. However, it should be noted that there are several distinct differences between the quantum walk parton shower and the quantum walks presented in Section~\ref{sec:speedup}, which are caveats to the potential speed up. The implementations presented in \cite{doi:10.1137/090745854} and \cite{Lemieux2020efficientquantum} extend Szegedy's quantum walk \cite{Szegedy} by employing quantum phase estimation to prepare steady states on the quantum device. Furthermore, it has only been proven that coin quantum walks can replicate Szegedy's quantisation of Markov Chains when the shift and coin operations are Hermitian~\cite{WongCoinedWalk, Portugal}, with other architectures still being an active field of research. With these caveats in mind, whether a speed up can be realised for the quantum walk parton shower warrants a further, in depth study. 

It is realistic to consider future extensions to the quantum walk parton shower algorithm, such as the introduction of colour, that would benefit from extensions such as phase estimation and amplitude amplification, thus maintaining or improving any potential speed up. A brief discussion about the extension of the simple parton shower example to a more realistic shower is given in Section~\ref{sec:future}.

\subsection{Towards a realistic parton shower}\label{sec:future}

The parton shower algorithm described in Section~\ref{sec:collinearShower} is a simplified, toy model and has thus limited capability compared to state-of-the-art, classical parton shower algorithms. However, the quantum algorithm leverages the unique ability of the quantum computer to remain in a superposition state throughout the calculation, enabling all shower histories to be calculated simultaneously and providing a remarkable advantage over the classical algorithms. 
It is interesting to consider how the parton shower algorithm will develop with advancements in quantum technologies. Near-future devices with larger quantum volume~\cite{gambetta_2020, Jurcevic2020DemonstrationOQ} make it feasible to imagine a practical parton shower algorithm on a quantum device.   

An obvious extension to the algorithm proposed would be to include more particle types and flavours. As described in Section~\ref{sec:collinearShower}, this is easily done by increasing the dimension of the $\mathcal{H}_P$ and $\mathcal{H}_C$ spaces to include another particle and its corresponding splittings. It may then be possible to extend the shower to include all quark flavours, increasing the dimension of the walker's lattice to seven: six quark dimensions and one gluon dimension. To implement this circuit would require a large number of qubits, with the number required for each particle type being

\begin{equation}
n_\textrm{qubits} = \log_2 N,
\end{equation}
where $N$ is the number of desired steps in the shower process. It is possible to reduce the overall number of qubits in the system by removing redundant areas in the quantum walker's lattice. For example, in Figures~\ref{fig:originalResults} and \ref{fig:modResults}, there is only one quark-antiquark pair in the results. Therefore, all lattice sites containing two or more quark-antiquark pairs could be removed to streamline the circuit. However, this does reduce the generality of the circuit, and such areas would have to be known \textit{a priori} to running the device.

The introduction of additional particles to the shower would eventually require the simulation to keep track of colour flow within the parton shower. In \textsc{Herwig}++~\cite{Herwig}, the colour reconnection stage of the parton shower is calculated using a classical simulated annealing process. It has been shown that quantum walk algorithms provide a quadratic speed up over classical simulated annealing~\cite{PhysRevA.78.042336, PhysRevLett.101.130504}, and is discussed further in Section~\ref{sec:speedup}. Consequently, it is expected that quantum walks will help improve the run time for more complex parton showers.

\begin{figure}[t]
\centering
\includegraphics[scale=0.4]{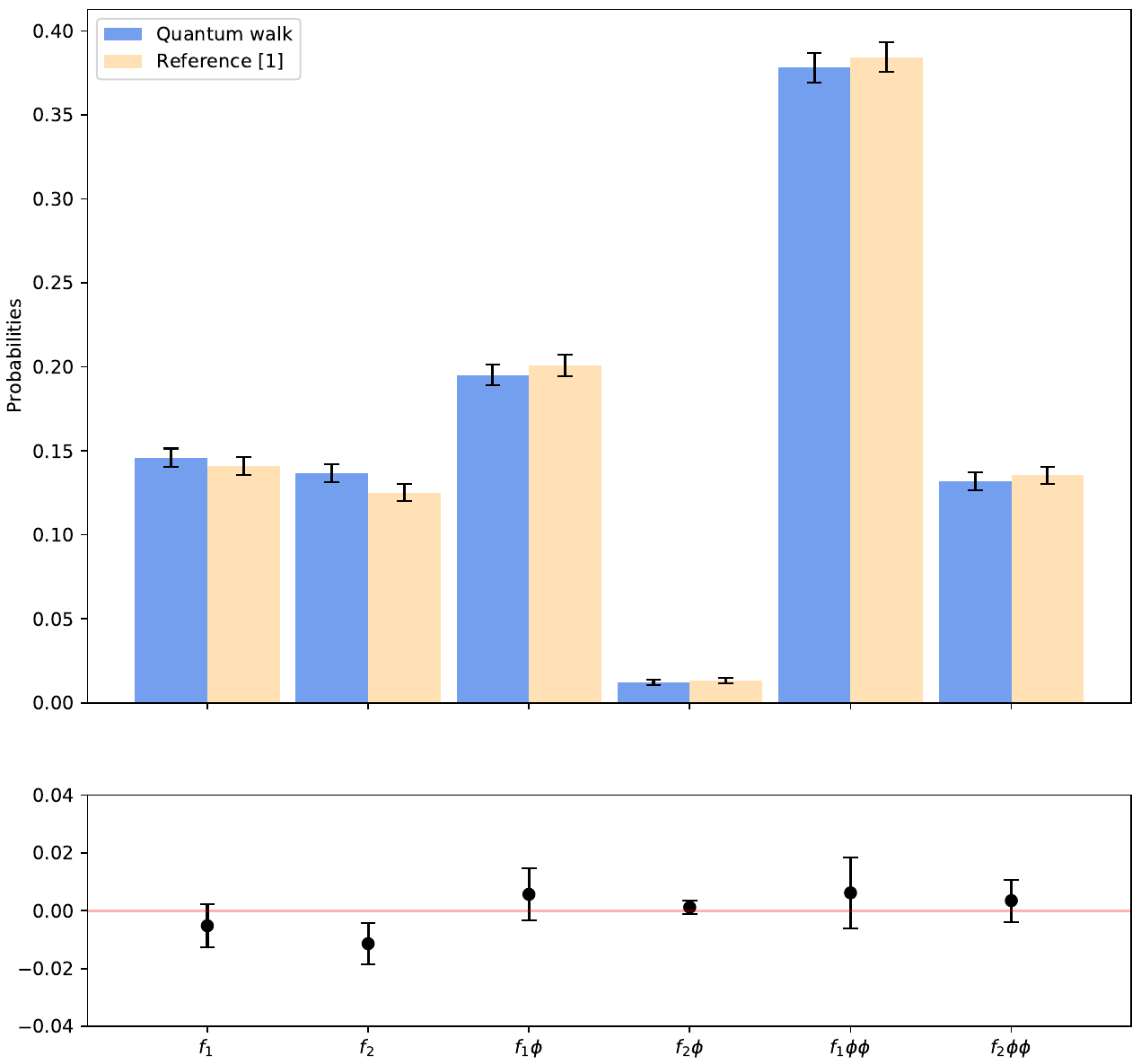}
\caption{Comparison between the quantum walk parton shower framework and the algorithm presented in Reference~\cite{bauer2019quantum}. Both algorithms have been run on the IBM Q 32-qubit quantum simulator for 5000 shots. There is good agreement between both algorithms, with the quantum walk framework offering an improvement in the required Quantum Volume. The quantum walk circuit is comprised of 6 qubits, an improvement on the 22 qubits required for the parton shower presented in~\cite{bauer2019quantum}.}
\label{fig:ref1Comp}
\end{figure}

It is feasible to consider an algorithm that can simulate a parton shower for calculations where a basis transformation is performed. For example, it has been argued in~\cite{bauer2019quantum} that a quantum advantage can be achieved by including interference effects into the sampling process of the final probability distribution. The authors present a special case consisting of two fermions, $f_1$ and $f_2$, and a scalar $\phi$. The algorithm performs a rotation from the flavour basis, $f_{1/2}$, to a mass basis, $f_{a/b}$. The parton shower calculation is then carried out in the mass basis, rotating back to the flavour basis before measurement. It is through this basis change that the authors find the interference effects are introduced and thus provides a quantum advantage over classical methods. Here, we have successfully replicated the 2-step parton shower from~\cite{bauer2019quantum} in the quantum walk framework (without $\phi$ splittings) by increasing the dimension of the Hilbert space to accommodate the two fermions and the scalar. Figure~\ref{fig:ref1Comp} shows a comparison between the results from both algorithms, and good agreement is obtained. The quantum walk parton shower is implemented on a circuit comprising 6 qubits, an improvement on the 22 qubits required for the parton shower presented in~\cite{bauer2019quantum}. Furthermore, the quantum walk parton shower has a shallow circuit depth, requiring only 61 gates (19 single qubit gates, 30 \textsc{ccnot} gates and 12 \textsc{cnot} gates) compared to 148 gates (45 single qubit gates, 74 \textsc{ccnot} gates and 29 \textsc{cnot} gates) in \cite{bauer2019quantum}. The basis transformation is applied across the fermion dimensions of the position space $\mathcal{H}_P$ and the quantum walk parton shower agrees well with the previous algorithm. Consequently, the quantum walk parton shower simulates the quantum interference effects and maintains the advantage claimed by~\cite{bauer2019quantum}.

\begin{figure}
\centering
\includegraphics[scale=0.45]{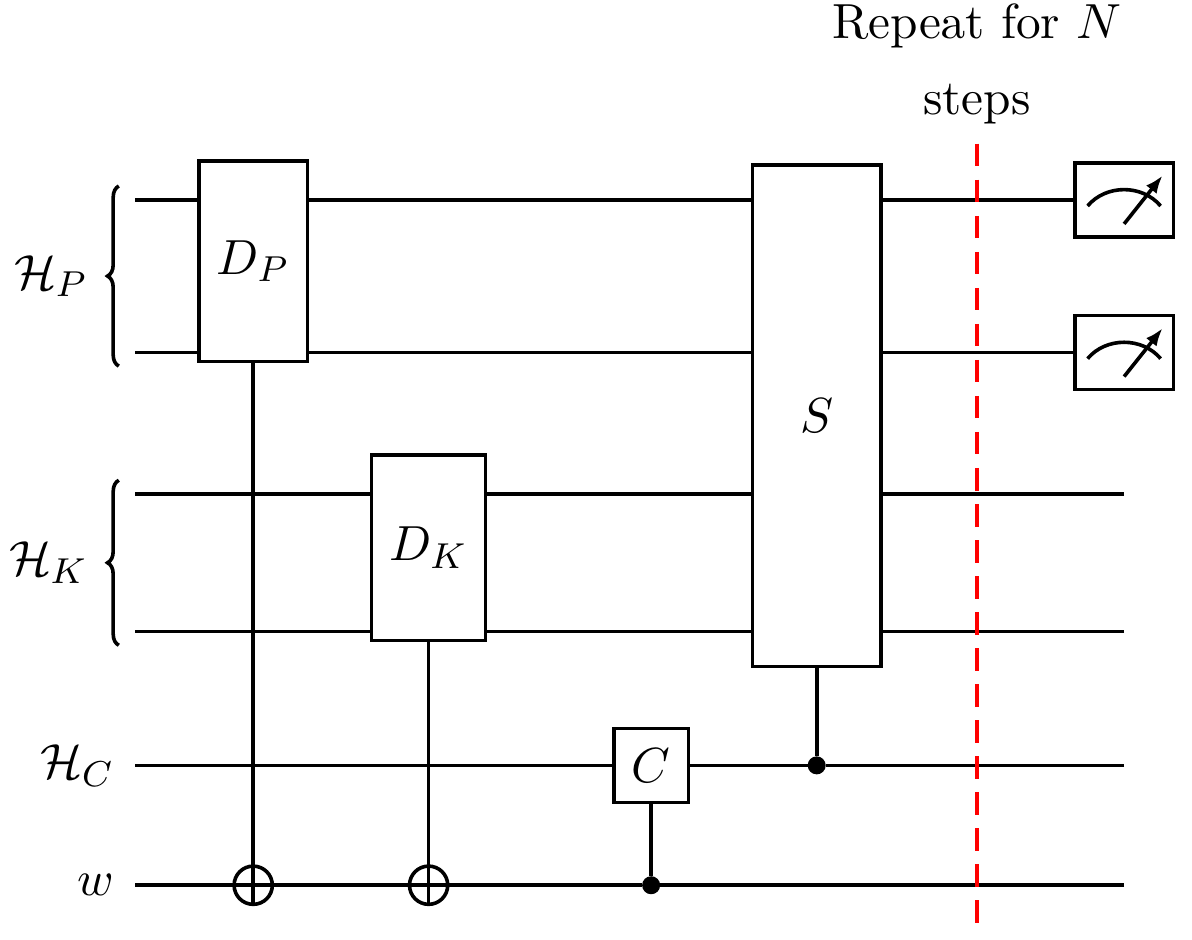}
\caption{A schematic circuit diagram for a one particle type parton shower with a discretised kinematic space. Here, $\mathcal{H}_P$, $\mathcal{H}_K$ and $\mathcal{H}_C$ are the position, kinematic and coin spaces respectively, and $w$ is an ancillary register.}
\label{fig:kinematics}
\end{figure}

Keeping track of particle kinematics in the parton shower algorithm outlined in Section~\ref{sec:collinearShower} is an important step towards emulating a realistic parton shower. The current publicly accessible devices and simulators do not have adequate quantum volume to include shower kinematics, but future devices may have the capability to implement this. Within the quantum walk framework, it is possible to consider extending the Hilbert space of the system to include a \textit{kinematic} space $\mathcal{H}_K$ such that the total space now has the form,

\begin{equation}
\mathcal{H} = \mathcal{H}_C \otimes \mathcal{H}_P \otimes \mathcal{H}_K.
\end{equation}
The kinematic space $\mathcal{H}_K$ would comprise a discretised momentum space that each shower particle could move in. Similarly to the position check in Figure~\ref{fig:partonShower}, conditional coin operations would then be used to apply the correct splitting kernels to the coin qubits depending on the position of the walker in the kinematic space $\mathcal{H}_K$. A schematic of a one particle type parton shower, with kinematics included, is shown in Figure~\ref{fig:kinematics}. It should be noted that, in order to keep track of each particle's momentum in the shower, the kinematic space $\mathcal{H}_K$ will have to be extended at each splitting. One can initialise the system to have the whole kinematic space at the beginning of the algorithm, populating the space only in the event of a splitting. However, this approach will lead to a large redundancy in the circuit, an area which may have to be optimised in practice. 

\section{Summary}\label{sec:Summary}

Simulating parton showers on quantum computers has been shown~\cite{bauer2019quantum, PhysRevD.103.076020} to have distinct advantages that exploit the unique features of the quantum device. 
In classical parton showers, all possible shower histories are calculated individually, stored on a physical memory device and then analysed in their entirety to provide information on a physical quantity. In contrast, the quantum device remains in a quantum state throughout the calculation, constructing a wavefunction which comprises a superposition of all possible shower histories. Consequently, all shower histories are calculated simultaneously in a single calculation, removing the requirement to store and track each shower history on physical memory. However, simply porting over the classical parton shower implementations onto a quantum device is computationally inefficient, requiring a large number of qubits and only allowing up to 2 steps of the parton shower to be simulated on current simulators~\cite{ PhysRevD.103.076020}. 

This paper proposes a novel quantum walk approach to simulating parton showers on a quantum computer that represents a significant improvement in the depths of the shower that can be simulated and with far fewer qubits. 
We present a quantum algorithm for the simulation of a collinear, 31-step parton shower implemented as a 2D quantum walk, where the coin flip represents the total parton emission probability, and the movement of the walker in the 2D space represents an emission corresponding to either gluons or a quark-antiquark pair.  Reframing the parton shower in this quantum walk paradigm enables a 31-step shower to be simulated, a dramatic improvement over previous quantum algorithms~\cite{bauer2019quantum, PhysRevD.103.076020}. The efficient implementation of the quantum walk allows for a smaller number of registers in the algorithm, which in turn grow, at most, logarithmically with the number of shower steps, as shown in Equation~\ref{eqn:scaling}. As a consequence, a 31-step shower can be run on the IBM Q 32-qubit quantum simulator~\cite{32_sim} with almost a factor of two reduction in the number of required qubits compared to a 2-step shower in the previous implementation~\cite{ PhysRevD.103.076020}. As a direct comparison, the quantum walk framework can recreate the shower presented in~\cite{PhysRevD.103.076020} using 9 qubits and 203 gate operations (59 single qubit gates, 98 \textsc{ccnot} gates and 46 \textsc{cnot} gates), compared to the 31 qubits and 444 gate operations (169 single qubit gates, 217 \textsc{ccnot} gates and 58 \textsc{cnot} gates) required in~\cite{PhysRevD.103.076020}. Furthermore, the algorithm has been shown to replicate the shower presented in~\cite{bauer2019quantum} using only 6 qubits and 61 gates (19 single qubit gates, 30 \textsc{ccnot} gates and 12 \textsc{cnot} gates), compared to 22 qubits and 148 gates (45 single qubit gates, 74 \textsc{ccnot} gates and 29 \textsc{cnot} gates) required in \cite{bauer2019quantum}. A comparison of the quantum walk parton shower and the shower presented in~\cite{bauer2019quantum} is shown in Figure~\ref{fig:ref1Comp}, and good agreement is obtained. The quantum walk approach thus offers a natural and much more efficient approach to simulating parton showers on quantum devices, and is a step towards simulating a realistic parton shower on a quantum computer.

\vspace{1cm}

\noindent {\it{{\bf Acknowledgements:}~~Sarah Malik and Simon Williams are funded by grants from the Royal Society. We would like to acknowledge the use of the IBM Q for this work. We thank Frank Krauss and Stefan Prestel for valuable discussions.}}

\bibliographystyle{JHEP}
\bibliography{references}

\end{document}